# *Nonradiating and radiating modes excited by quantum emitters in open epsilon-near-zero cavities*

*Iñigo Liberal[1] and Nader Engheta[1*]*

[1]Department of Electrical and Systems Engineering

University of Pennsylvania

Philadelphia, Pennsylvania 19104, USA

*Correspondence to: engheta@ee.upenn.edu

## Abstract

Controlling the emission and interaction properties of quantum emitters (QEs) embedded within an optical cavity is a key technique in engineering light-matter interactions at the nanoscale, as well as in the development of quantum information processing. State-of-the-art optical cavities are based on high Q photonics crystals and dielectric resonators. However, wealthier responses might be attainable with cavities carved in more exotic materials. Here, we theoretically investigate the emission and interaction properties of QEs embedded in open epsilon-near-zero (ENZ) cavities. Using analytical methods and numerical simulations, it is demonstrated that open ENZ cavities present the unique property of supporting nonradiating modes independently of the geometry of the external boundary of the cavity (shape, size, topology…). Moreover, the possibility of switching between radiating and nonradiating modes enables a dynamic control of both the emission by, and the interaction between, QEs. These phenomena provide unprecedented degrees of freedom in controlling and trapping fields within optical cavities, as well as in the design of cavity opto- and acousto-mechanical systems.

Cavity quantum electrodynamics (cavity QED) is the field of research that investigates the interaction between quantum emitters (QEs), such as atoms and quantum dots, and a resonant cavity [1]. This interaction is fundamentally interesting, and it could be the basis for quantum information processing [2]. The QE-cavity interaction is also relevant for single-photon sources [3], single-photon nonlinearities [4], lasing [5,6] and quantum many body systems [7,8]. Owing to their low losses and associated high quality factors, cavities constructed using photonics crystals [9–13], optical microcavities [14] and Anderson-localized modes [15] are commonly employed. Despite losses, plasmonic systems [16–19] are also attractive as they provide subwavelength confinements, with cavity sizes well below the diffraction limit.

Aside from this spectrum of conventional cavities, more sophisticated responses in the emission and interaction properties of QEs could be obtained with cavities carved in more exotic materials. For instance, zero-index metamaterials (e.g., epsilon-near-zero (ENZ) [20] or epsilon-mu-near-zero (EMNZ) media [21–23]) exhibit a decoupling between spatial and temporal field variations [21,24], which enables numerous wave phenomena including: tunneling [20,25], geometry-invariant eigenfrequencies [26], electric levitation [27], and unconventional force density distributions [28]. In terms of tailoring the emission properties of QEs, the phase uniformity in zero-index metamaterials has been exploited in order to enhance the directivity of single emitters [29,30], as well as to construct collective interference effects among multiple emitters [23,30–32]. The wealth in wave phenomena related to metamaterials with near-zero parameters, as well as their potentiality in enhancing emission properties, has motivated us to investigate the emission properties of QEs embedded in open ENZ cavities, with a view towards their future application in cavity QEDs. Note that far from being a theoretical curiosity, there are several experimental demonstrations of zero-index metamaterials based on naturally available materials [33,34], dispersion engineering in waveguides [25,35], photonic crystals [36] and artificial electromagnetic materials [37,38].

We demonstrate that the main signature of a QE embedded in an open ENZ cavity is the excitation of a nonradiating mode independently of the geometry of the external boundary of the cavity (shape, size, topology…). Nonradiating modes have been investigated for a long time due to their connection to classical problems such as models for stable atoms and elementary particles (see, e.g., [39,40]). Furthermore, the extreme light confinement facilitated by nonradiating modes may also have practical applications in nonlinear optics, sensing, and heating [41,42], the storage of 'bits' of quantized energy light [43], as well as in managing the reactive power surrounding an emitter [44]. Recently, the excitation of nonradiating modes in spherical plasmonic cavities has been investigated [41–43,45]. Here, we demonstrate theoretically that these apparently exotic nonradiating modes can actually be excited in open cavities of arbitrary geometry, and that they provide unique opportunities in controlling the emission properties of QEs.

We start by considering the emission properties of a QE located at the center of a spherical vacuum bubble of radius $r_0$, which is itself immersed in an unbounded ENZ environment (see Fig. 1a). The insulating bubble is required to avoid the singularity that arises from the direct contact of a source with an absorbing medium [46,47]. The QE is modeled as a point-like two-level system with dipole moment $\boldsymbol{p} = p\,\hat{\boldsymbol{z}}$, intrinsic nonradiative decay rate $\Gamma_{NR}$, and transition frequency $\omega_0$ [48], which is assumed to be centered at the ENZ frequency of the background medium ($\omega_0 \approx \omega_p$). Moreover, we set $\omega_0 \approx \omega_p \approx 2\pi \times 29.08 \times 10^{12}$ rad/s (i.e., $\lambda_0 \approx \lambda_p \approx 10.32\ \mu$m) corresponding to the plasma frequency of silicon carbide (SiC) [33] to facilitate future experimental demonstrations. Time-harmonic field expressions $\exp(-i\omega t)$ are assumed and omitted hereafter. It can be shown (supplementary material, sections 3.1 and 3.2) that the *spatial distribution* of the classical fields excited by the aforementioned insulated dipole immersed in an unbounded ENZ medium are equal to those of an electrostatic dipole (even though the fields oscillate in time at frequency $\omega_0$), with effective dipole moment $p_{\text{eff}}$, i.e.:

$$\boldsymbol{E}(\boldsymbol{r}) = \frac{1}{4\pi\varepsilon_0} \frac{\hat{\boldsymbol{r}}\, 2\cos\theta + \hat{\boldsymbol{\theta}}\, \sin\theta}{r^3}\, p_{\text{eff}} \tag{1}$$

$$\boldsymbol{H}(\boldsymbol{r}) = \boldsymbol{0} \tag{2}$$

Intriguingly, this implies that the magnetic field in the ENZ region is zero, and it is indeed trapped within the vacuum bubble. This fact can be clearly appreciated in Fig. 1b, which represents the electric and magnetic field magnitude distributions as computed numerically for a QE embedded in a spherical vacuum bubble of radius $r_0$ = 0.25 µm [49]. Therefore, we find that insulated QEs embedded in ENZ media can be effectively treated as sources of spatially electrostatic fields with a time-harmonic variation. The strength of the effective electrostatic dipole moment $p_{\text{eff}}$ is determined by the properties of the vacuum bubble (supplementary material, section 3.2):

$$p_{\text{eff}} = \frac{(k_0 r_0)^2}{\hat{J}_1(k_0 r_0)}\, p \tag{3}$$

where $k_0 = \omega_0/c$ and $\hat{J}_1(x) \equiv \sqrt{\frac{\pi x}{2}}\, J_{\frac{3}{2}}(x)$ is the Schelkunoff form of the spherical Bessel functions of the first kind and order one, where $J_n(x)$ is the cylindrical Bessel function of the first kind and order one [50]. For example, the effective dipole moment is three times the original dipole moment for deeply subwavelength bubbles, i.e., $k_0 r_0 \ll 1 \to p_{\text{eff}} \approx 3\, p$, in consistence with the quasi-static solution to the problem (supplementary material, section 3.3). Furthermore, it is apparent from (3) that the cavity conformed by the vacuum bubble becomes resonant at the zeros of $\hat{J}_1(k_0 r_0)$. Moreover, since the spatially electrostatic nature of the fields (and the lack of magnetic field in the ideal ENZ region) prevents radiation losses, arbitrarily large effective dipoles can be excited at resonance in the absence of dissipation losses. Naturally, the magnitude of $p_{\text{eff}}$ at resonance (i.e., at $\hat{J}_1(k_0 r_0) = 0$) is ultimately limited in practice by dissipation losses and, specifically, it can be approximated by (supplementary material, section 3.4)

$$p_{\text{eff}} \approx -\frac{i}{\varepsilon''} \frac{(k_0 r_0)}{\hat{J}_1'(k_0 r_0)} p \approx -i \frac{4.6}{\varepsilon''} p \qquad (4)$$

where $\hat{J}_1'(x) = \partial_x \{\hat{J}_1(x)\}$ and $\varepsilon''$ is the imaginary part of the relative permittivity of the ENZ host medium. This simple analytical rule is validated in Fig. 1c, which depicts the effective dipole moment as computed with a full-wave numerical solver [49]. Note that even with an amount of loss comparable to that of naturally available ENZ materials ($\varepsilon''$= 0.1 [33]), the magnitude of the effective dipole is 46 times larger than that of the original dipole. Equivalently, the electric field intensity in the vicinity of the bubble is enhanced by a factor of $(46)^2 = 2116$, which suggests applications in enhancing the coupling between the QE and its environment, as well as in triggering nonlinear phenomena.

Next, as illustrated in Fig. 1d (see also Fig. S1), the QE is placed within an ENZ environment of finite size, conforming an open cavity. In the most general case, the cavity can be of any geometry, including the presence of other dielectric bodies within it. In this generalized scenario, the fields excited in the ENZ host will be those of the unbounded case, plus the fields "scattered" at the interface of the cavity with the unbounded external space and the dielectric bodies. Since the sources of the problem initially excite spatially electrostatic fields within the ENZ region, a valid solution to the scattering problem is the spatial distribution determined by the solution of the spatially electrostatic problem. Owing to the uniqueness of the solution, this is indeed the spatial distribution of the fields excited in the time-harmonic case. In this manner, the electric field in the ENZ region can be written as the gradient of a scalar potential $\boldsymbol{E} = -\nabla V$, and, consequently, it corresponds to the solution of the Laplace equation $\nabla^2 V = 0$ subject to the boundary conditions imposed by the continuity of both $V$ and $\varepsilon\, \partial V/\partial n$. As noted in [51], the Laplace equation is independent of the background permittivity, which only appears in the solution through the boundary conditions. Interestingly, only the ratio between the permittivities at each side of the interface matters. Thus, the field distribution of a body of permittivity $\varepsilon_2$ immersed in a background of permittivity $\varepsilon_1$, is identical to that of a body of permittivity $\varepsilon_2/\varepsilon_1$ immersed in vacuum [51]. Therefore, when the background medium is ENZ, $\varepsilon_1 \approx 0$, all material bodies behave as effectively perfect electric conductors, $\varepsilon_2/\varepsilon_1 \to \infty$, for spatially electrostatic fields. Consequently, the fields excited by the QE will be unable either to escape the cavity or to penetrate any dielectric body within this ENZ region. This effect is clearly illustrated in Fig. 1e, which depicts the electric and magnetic field magnitude distributions [49]. The magnetic field is again trapped within the vacuum bubble containing the QE at its center. By contrast, the electric field penetrates within the ENZ cavity in the form of a time-varying electrostatic field, but it is nonetheless unable to either escape the cavity or enter the other dielectric bodies.

We emphasize that this property is independent of the geometry of the external boundary of the cavity and/or the dielectric bodies contained within it. This geometry-invariant confinement can also be understood by noting that the bound charges excited at the interface of ENZ and vacuum has distributions identical to those that would be excited at the interface of vacuum and a perfect electrical conductor (PEC) of analogous geometry. The former are given by $\sigma_b = -\nabla \cdot \boldsymbol{P} = \varepsilon_0(\varepsilon - 1)E_{n1} = -\varepsilon_0 E_{n1}$, whereas the latter would be given by $\sigma_f = \nabla \cdot \boldsymbol{D} = -\varepsilon_0 E_{n1}$, where $E_{n1}$ is the electric field normal to the interface in the first medium. Since identical charge distributions give rise to the same electric field, it is clear that, in terms of finding the solution to the "scattered" field, all boundaries in contact to the ENZ medium effectively behave as PEC boundaries. In this manner, it is demonstrated that insulated QEs embedded in open ENZ cavities excite nonradiating modes that are confined within the extent of the cavity independently of its geometry, and even when the cavity itself is open to an unbounded vacuum space. We emphasize that although nonradiating modes have already been predicted in open spherical plasmonic cavities [41,42,45], here we demonstrate that these modes exist in open ENZ cavities *independently of the geometry of its external boundary*, and that this effect is empowered by the spatially electrostatic nature of the fields. Moreover, the synergy between the confinement properties of nonradiating modes and the field intensity enhancements at the resonance of the effective electrostatic dipole enable high intensity localized fields on a volume prescribed on demand by the geometry of the cavity.

The above analysis is valid as long as the fields excited by the QE-vacuum bubble system are dominated by the electric dipole mode in the bubble. This is exactly the case when the QE is at the center of a spherical bubble, and it is a very accurate estimation for subwavelength bubbles. At the same time, bubbles with larger sizes can efficiently excite different, possibly radiating, modes. Far from being a limitation, this opens up the possibility of switching between nonradiating and radiating modes, and hence activating/deactivating the interaction of QEs with a system external to the cavity. For example, and as schematically depicted in Fig. 2a (see also Fig. S2), one can take an open ENZ cavity with arbitrarily shaped external boundary containing a spherical vacuum bubble whose radius has been tuned to be resonant with a radiating mode (e.g., a magnetic dipolar mode). However, due to the symmetry of the fields, only the electric dipolar mode is excited when the QE is at the center of the sphere, resulting in a nonradiating cavity mode as mentioned above. However, as the QE is shifted from the center of the bubble, its field can be decomposed as a series of multipoles [52], and the QE excites the resonant radiating mode. In practice, the position of the QE can be controlled with different mechanisms, e.g., sound waves, microelectromechanical systems (MEMS), optomechanical techniques, etc. One of these options is schematically illustrated in Fig. 2a, in which the QE is assumed to be attached to a membrane in the cavity. Thus, if the cavity were excited by an external optical or acoustic wave, the membrane would vibrate and the position of the QE and, hence, its emission properties, would oscillate in synchrony with the vibrational mode of the cavity.

This effect can be appreciated in Fig. 2b which depicts the simulated electric field distribution for the QE emitter at two different emitter positions [49]. As anticipated, a nonradiating mode is excited for $\Delta x = 0$, and the field is confined within the cavity. By contrast, the field is strongly radiated outside the cavity for $\Delta x = 3.5\,\mu$m. In our numerical simulation, the cavity has been considered with an arbitrary (not particularly designed) non-canonical external boundary, although the radius of the internal spherical vacuum bubble has been numerically optimized to trigger the magnetic dipolar resonance when the QE is displaced away from the bubble's center. The resulting optimal radius, $r_0$ = 5.27 μm, is indeed close to the value of the radius that triggers the magnetic dipolar resonance for a vacuum sphere immersed in unbounded ENZ medium (supplementary material, section 3.5). In practice, the geometry of the external boundary of the cavity could be engineered for different purposes. These include, among others, boosting the emission of the radiating mode, improving the coupling with an optical or acoustic wave and/or exciting specific vibrational modes, catalyzing the interaction with light at other frequencies where the cavity is transparent, and/or just providing the intriguing possibility of having a deformable device.

We make use of the nanoantenna formalism developed in [53,54] in order to quantitatively assess the emission properties of a QE embedded in this specific cavity. In particular, we calculate the normalized excitation rate $\Gamma_{\text{exc}}/\Gamma_{\text{exc}}^{\text{free}}$ (i.e., the rate of excitation via spontaneous emission of a receiver located outside the cavity, at the position schematically depicted in Fig. 2a, and normalized to the free-space excitation rate) and the quantum yield $\eta_{\text{rad}}$ (or radiation efficiency, i.e., the ratio of radiative to total decay rates, where the latter includes both the dissipation decay in the cavity and the intrinsic nonradiative decay rate) [53,54]. Both quantities are depicted in Figs. 2c and 2d as a function of the emitter position displacement, $\Delta x$, and for different amounts of loss of the ENZ host medium $\varepsilon''$. For illustrative purposes, we assume that the intrinsic quantum yield is 0.5. It is evident from Figs. 2c and 2d that both the excitation rate and the radiation efficiency are suppressed when the dipole is at the center of the vacuum bubble, consistent with the excitation of non-radiating modes. Moreover, both figures of merit increase as the QE is shifted from the center, and they reach a maximum at the specific displacement $\Delta x = 3.5 \ \mu$m. This optimal value corresponds to the maxima of the magnetic dipolar coefficient as computed from the addition theorem (supplementary material, section 3.6). Naturally, the excitation rate and radiation efficiency are limited by the losses of the ENZ medium. However, we note that even with high losses $\varepsilon'' = 0.1$ the excitation rate is enhanced by a factor of five with respect to that of free-space (c.f., Fig. 2c). These results indicate that effect of switching between radiating and non-radiating modes could be measured in experimental setups based on naturally available ENZ materials, though better performances could be obtained with synthetic ENZ media (such as waveguides at cut-off frequencies [25,35]). Furthermore, we emphasize that a similar effect takes place for near-field interactions since, as shown in Fig. 2b, the near field outside the cavity is also suppressed when a nonradiating mode is excited.

In this manner, the mechanism of switching between nonradiating and radiating modes also empowers a dynamic control of the dipole-dipole interactions between different emitters. As depicted in Fig. 3a, one could embed two emitters, e.g., $p_1$ and $p_2$, in different spherical bubbles placed within an open ENZ cavity. When both emitters are at the center of their bubbles they excite nonradiating modes and they are effectively decoupled. However, they become resonantly coupled as their position separates from such a symmetric position. In this case, we can assume that the position of the QEs may be controlled by means, e.g., MEMS placed outside the cavity. To this end, the cavity may be again assumed to be formed by an ENZ host with a noncanonical geometry (see Figs. 3a and S3), whereas spherical bubbles can be assumed to be made of silicon (Si), characterized by relative permittivity $\varepsilon_{Si} = 11.7$, and their radius $r_0$ = 1.505 µm has been tuned to trigger the magnetic dipolar resonance (supplementary material, section 3.5). In order to facilitate the control of the QEs position via, e.g., a MEMS system, perhaps the cavity may be pierced by a silicon rod of section 0.25 *µ*m x 0.25 *µ*m. Thus, this example also serves to illustrate that the proposed cavities would be robust against the modifications that could be required to implement them in practice (e.g., to include a Si rod piercing the cavity).

The simulation results for the coupling/decoupling mechanism mediated by the nonradiating and radiating modes is clearly illustrated in Fig. 3b, which represents the electric field magnitude distributions excited by the QE in the left bubble for displacements $\Delta x = 0$ and $\Delta x = 1\,\mu m$ [49]. Note that in our simulation the presence of the Si rod has no appreciable impact on the radiating and nonradiating nature of the fields. We also make use of dyadic Green's function-based field quantization scheme in order to quantitatively estimate the performance of this coupling/decoupling mechanism [55]. Specifically, we compute the contribution to the decay rate related to the coupling between both emitters, $\Gamma_{21} = \frac{2k_0^2}{\hbar\varepsilon_0} \boldsymbol{p}_2 \cdot \text{Im}\{\overline{\overline{\boldsymbol{G}}}(\boldsymbol{r}_2, \boldsymbol{r}_1)\} \cdot \boldsymbol{p}_1$, as well as the photonic Lamb shift produced by this interaction, $\Delta\omega_{21} = -\frac{k_0^2}{\hbar\varepsilon_0} \boldsymbol{p}_2 \cdot \text{Re}\{\overline{\overline{\boldsymbol{G}}}(\boldsymbol{r}_2, \boldsymbol{r}_1)\} \cdot \boldsymbol{p}_1$ [56]. Here, $\overline{\overline{\boldsymbol{G}}}(\boldsymbol{r}_2, \boldsymbol{r}_1)$ is the dyadic Green's function describing the field excitation at $\boldsymbol{r}_2$ produced by the emitter located at position $\boldsymbol{r}_1$. This function was numerically evaluated and $\Gamma_{21}$ and $\Delta\omega_{21}$ are depicted in Figs. 3c and 3d, respectively, normalized with respect to their free-space counterparts. It is evident from these figures that both dipoles are effectively decoupled when they are at the center of their respective silicon spheres $\Delta x = 0$ μm. However, the coupling increases as the emitters are shifted away from such a symmetric configuration and, in accordance to the addition theorem (supplementary material, section 3.6), it is optimized for $\Delta x = 1$ μm. Note that while the strength of the coupling depends on the losses of the ENZ cavity, it is found that even with the relatively high losses $\varepsilon'' = 0.1$ the coupling still exhibits significant enhancements, $|\Gamma_{21}/\Gamma_{21}^{\text{free}}| \sim 113$ and $|\Delta\omega_{21}/\Delta\omega_{21}^{\text{free}}| \sim 27$, with respect to the free-space case. Moreover, we note that this geometry could be straightforwardly extended to a multi-emitter system by adding more vacuum bubbles containing emitters, which suggests interesting applications in recreating many-body quantum problems.

Our results demonstrate that bubble-insulated QEs embedded in open ENZ cavities present the unique signature of exciting nonradiating modes independently of the geometry of the external boundary of the cavity. This effect provides unprecedented degrees of freedom in controlling and trapping electromagnetic fields within an open optical cavity. In addition, our study reveals that it is possible to switch between nonradiating and radiating modes, providing new venues in controlling the emission properties of QEs, such as enhancing/suppressing the spontaneous emission exiting the cavity, as well as dynamically activating/deactivating the coupling between QEs. The fact that these effects can take place in cavities with arbitrarily shaped boundaries could be exploited to resonantly couple with other physical processes, such as sound waves, enabling the coupling with specifically designed cavity-induced vibrational modes. The geometry could also be tailored to boost the emission of radiating modes, and/or to facilitate the excitation and manipulation of QEs with electromagnetic waves operating at frequencies where the cavity could be transparent or resonant.

This work is supported in part by the US Air Force Office of Scientific Research (AFOSR) Multidisciplinary University Research Initiatives (MURI) on Quantum Metaphotonics & Metamaterials, Award No. FA9550-12-1-0488

**References:**

[1]   P. R. Berman, *Cavity Quantum Electrodynamics* (Academic Press, Boston, 1994).

[2]   H. J. Kimble, Nature **453**, 1023 (2008).

[3]   J. Claudon, J. Bleuse, N. S. Malik, M. Bazin, P. Jaffrennou, N. Gregersen, C. Sauvan, P. Lalanne, and J.-M. Gérard, Nat. Photonics **4**, 174 (2010).

[4]   D. E. Chang, A. S. Sørensen, E. A. Demler, and M. D. Lukin, Nat. Phys. **3**, 807 (2007).

[5]   J. McKeever, A. Boca, A. D. Boozer, J. R. Buck, and H. J. Kimble, Nature **425**, 268 (2003).

[6]   M. A. Noginov, G. Zhu, A. M. Belgrave, R. Bakker, V. M. Shalaev, E. E. Narimanov, S. Stout, E. Herz, T. Suteewong, and U. Wiesner, Nature **460**, 1110 (2009).

[7] A. D. Greentree, C. Tahan, J. H. Cole, and L. C. L. Hollenberg, Nat. Phys. **2**, 856 (2006).

[8] M. J. Hartmann, F. G. S. L. Brandão, and M. B. Plenio, Nat. Phys. **2**, 846 (2006).

[9] T. Yoshie, A. Scherer, J. Hendrickson, G. Khitrova, H. M. Gibbs, G. Rupper, C. Ell, O. B. Shchekin, and D. G. Deppe, Nature **432**, 200 (2004).

[10] K. Hennessy, A. Badolato, M. Winger, D. Gerace, M. Atature, S. Gulde, S. Falt, E. L. Hu, and A. Imamoglu, Nature **445**, 14 (2006).

[11] M. Arcari, I. Söllner, A. Javadi, S. L. Hansen, S. Mahmoodian, J. Liu, H. Thyrrestrup, E. H. Lee, J. D. Song, S. Stobbe, and P. Lodahl, Phys. Rev. Lett. **093603**, 1 (2014).

[12] A. González-Tudela, C. L. Hung, D. E. Chang, J. I. Cirac, and H. J. Kimble, Nat. Photonics **1**, 16 (2015).

[13] J. S. Douglas, H. Habibian, C.-L. Hung, a. V. Gorshkov, H. J. Kimble, and D. E. Chang, Nat. Photonics **9**, 326 (2015).

[14] K. J. Vahala, Nature **424**, 839 (2005).

[15] L. Sapienza, H. Thyrrestrup, S. Stobbe, P. D. Garcia, S. Smolka, and P. Lodahl, Science **327**, 1352 (2010).

[16] M. Gullans, T. G. Tiecke, D. E. Chang, J. Feist, J. D. Thompson, J. I. Cirac, P. Zoller, and M. D. Lukin, Phys. Rev. Lett. **109**, 1 (2012).

[17] M. S. Tame, K. R. McEnery, S. K. Ozdemir, J. Lee, S. A. Maier, and M. S. Kim, Nat. Phys. **9**, 329 (2013).

[18] A. Delga, J. Feist, J. Bravo-Abad, and F. J. Garcia-Vidal, Phys. Rev. Lett. **112**, 1 (2014).

[19] S. Savasta, R. Saija, A. Ridolfo, O. Di Stefano, P. Denti, and F. Borghese, ACS Nano **4**, 6369 (2010).

[20] M. G. Silveirinha and N. Engheta, Phys. Rev. Lett. **97**, 157403 (2006).

[21]  R. W. Ziolkowski, Phys. Rev. E **70**, 046608 (2004).

[22]  N. Engheta and R. W. Ziolkowski, *Metamaterials: Physics and Engineering Explorations* (2006).

[23]  A. M. Mahmoud and N. Engheta, Nat. Commun. **5**, 5638 (2014).

[24]  N. Engheta, Science **340**, 286 (2013).

[25]  B. Edwards, A. Alù, M. E. Young, M. G. Silveirinha, and N. Engheta, Phys. Rev. Lett. **100**, (2008).

[26]  I. Liberal, A. M. Mahmoud, and N. Engheta, arXiv:1510.07005v1 (2015).

[27]  F. J. Rodríguez-Fortuño, A. Vakil, and N. Engheta, Phys. Rev. Lett. **112**, 033902 (2014).

[28]  I. Liberal, I. Ederra, R. Gonzalo, and R. W. Ziolkowski, Phys. Rev. A **88**, 053808 (2013).

[29]  S. Enoch, G. Tayeb, P. Sabouroux, N. Guérin, and P. Vincent, Phys. Rev. Lett. **89**, 213902 (2002).

[30]  A. Alù, M. G. Silveirinha, A. Salandrino, and N. Engheta, Phys. Rev. B **75**, 155410 (2007).

[31]  R. Fleury and A. Alù, Phys. Rev. B **87**, 201101 (2013).

[32]  J. J. Yang, Y. Francescato, S. A. Maier, F. Mao, and M. Huang, Opt. Express **22**, 9107 (2014).

[33]  W. G. Spitzer, D. Kleinman, and D. Walsh, Phys. Rev. **113**, 127 (1959).

[34]  G. V. Naik, J. Kim, and A. Boltasseva, Opt. Mater. Express **1**, 1090 (2011).

[35]  E. J. R. Vesseur, T. Coenen, H. Caglayan, N. Engheta, and A. Polman, Phys. Rev. Lett. **110**, 1 (2013).

[36]  X. Huang, Y. Lai, Z. H. Hang, H. Zheng, and C. T. Chan, Nat. Mater. **10**, 582 (2011).

[37]  R. Maas, J. Parsons, N. Engheta, and A. Polman, Nat. Photonics **7**, 907 (2013).

[38]  C. Rizza, A. Di Falco, and A. Ciattoni, Appl. Phys. Lett. **99**, (2011).

[39]  A. J. Devaney and E. Wolf, Phys. Rev. D **8**, 1044 (1973).

[40]  E. A. Marengo and R. W. Ziolkowski, Phys. Rev. Lett. **83**, 3345 (1999).

[41]  M. G. Silveirinha, Phys. Rev. A **89**, 023813 (2014).

[42]  F. Monticone and A. Alù, Phys. Rev. Lett. **112**, 1 (2014).

[43]  S. Lannebère and M. G. Silveirinha, Nat. Commun. **6**, 8766 (2015).

[44]  E. A. Marengo and R. W. Ziolkowski, IEEE Trans. Antennas Propag. **48**, 1553 (2000).

[45]  A. Erentok and R. W. Ziolkowski, IEEE Trans. Antennas Propag. **55**, 731 (2007).

[46]  S. Scheel, L. Knoll, and D.-G. Welsch, Phys. Rev. A **60**, 4094 (1999).

[47]  C. T. Tai and R. E. Collin, IEEE Trans. Antennas Propag. **48**, 1501 (2000).

[48]  L. Novotny and B. Hech, *Principles of Nano-Optics* (Cambrigde University Press, Cambridge, 2006).

[49]  *The Simulations Were Carried out in Comsol Multiphysics 5.0. See Supplementary Material for a Detailed Description of the Numerical Simulation Setup.* (n.d.).

[50]  R. F. Harrington, *Time-Harmonic Electromagnetic Fields* (McGraw-Hill, New York, 1961).

[51]  L. D. Landau and E. M. Lifshitz, *Electrodynamics of Continuous Media* (Pergamon, Oxford, UK, 1960).

[52] I. Liberal, I. Ederra, R. Gonzalo, and R. W. Ziolkowski, IEEE Trans. Antennas Propag. **61**, 5184 (2013).

[53] L. Novotny and N. van Hulst, Nat. Photon. **5**, 83 (2011).

[54] P. Bharadwaj, B. Deutsch, and L. Novotny, Adv. Opt. Photonics **1**, 438 (2009).

[55] P. Yao, C. Van Vlack, A. Reza, M. Patterson, M. M. Dignam, and S. Hughes, Phys. Rev. B **80**, 1 (2009).

[56] P. A. Huidobro, A. Y. Nikitin, C. González-Ballestero, L. Martín-Moreno, and F. J. García-Vidal, Phys. Rev. B **85**, 155438 (2012).

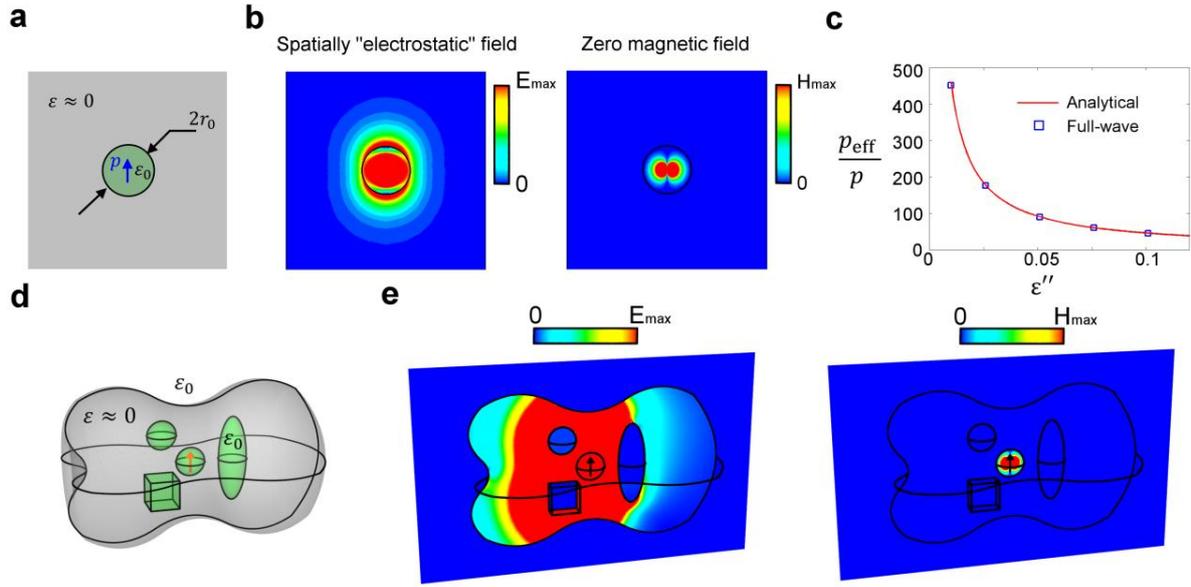

**Fig. 1. a,** Geometry and sketch of an QE with dipole moment $\boldsymbol{p}$ at transition frequency $\omega_0$ located at the center of a vacuum spherical bubble of radius $r_0$, embedded in an unbounded ENZ medium ($\varepsilon(\omega_0) \approx 0$). **b,** Simulated electric and magnetic field magnitude distributions ($r_0$=0.25 µm, $\lambda_0$= $\lambda_p$=10.31 µm). **c,** Analytically and numerically computed effective dipole moment enhancement factor at resonance ($r_0/\lambda_0 = 0.715$), as a function of losses (imaginary part of the permittivity of the ENZ region). **d,** Sketch of a bubble-insulated QE (shown as red arrow) embedded within an *open* ENZ cavity of arbitrary shape (shown as grey background) with several vacuum bubbles (shown in green). **e,f,** Simulated electric and magnetic field magnitude distributions. The electric field is trapped within the open cavity, while the magnetic field is confined to the vacuum bubble.

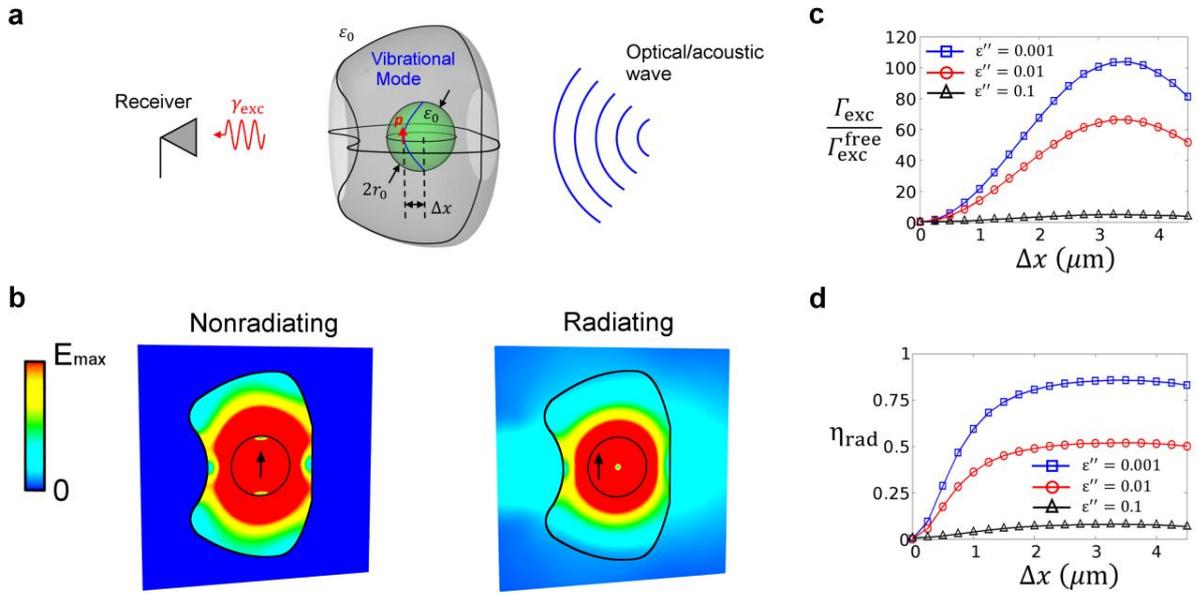

**Fig. 2. a,** Geometry and sketch of an arbitrarily shaped *open* ENZ cavity (shown as grey background), with a vacuum spherical bubble (shown in green, $r_0$ = 5.27 μm), containing a QE (shown as red arrow) attached to a membrane (shown as blue line), so that its position may be displaced along the x-axis due to external stimulus, e.g., the vibrational modes excited by an external optical/acoustic wave. **b,** Simulated (at $\lambda_0=\lambda_p$=10.31 μm) electric field magnitude distributions for displacements: *Δx = 0* (nonradiating mode) and *Δx = 3.5 μm* (radiating mode). **c,** Excitation rate $\Gamma_{exc}$ (normalized to the free space excitation rate) and **d,** quantum yield $\eta_{rad}$, as a function of emitter displacement, for different amounts of loss in the ENZ medium ($\varepsilon''$) and assuming an intrinsic quantum yield $\eta_{int}$ = 0.5.

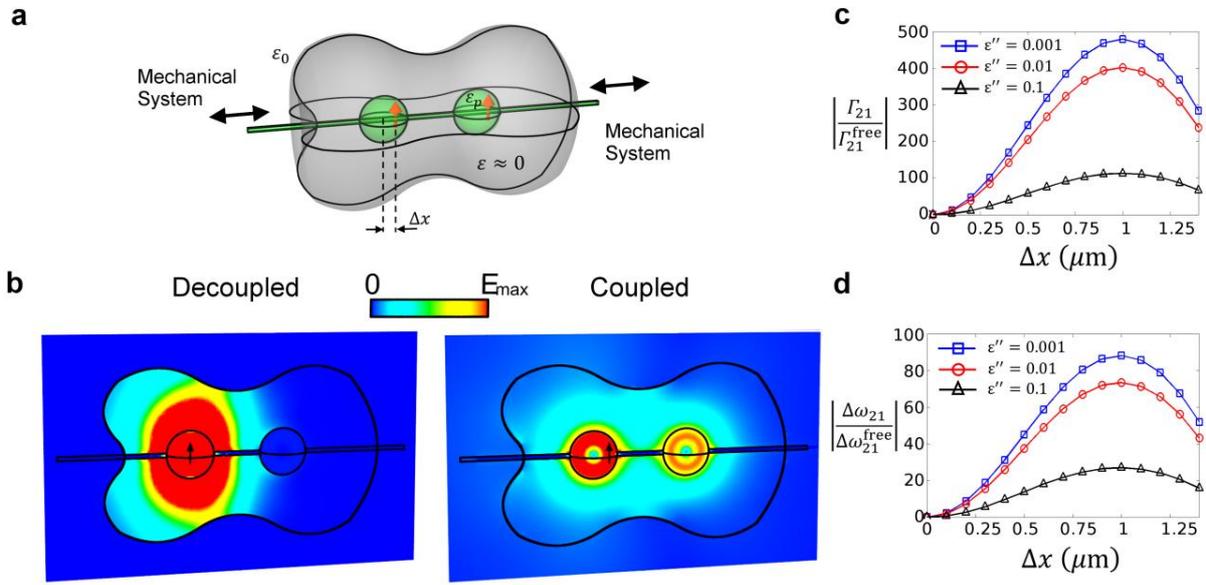

**Fig. 3. a,** Geometry and sketch of an *open* ENZ cavity of arbitrary shape (shown as grey background) containing two Si spherical bubbles (shown in green, $r_0$ = *1.505 μ*m, $\varepsilon_{Si}$ = 11.7) containing QEs (shown as orange arrows). The cavity is assumed to be pierced by a Si rod of cross-section 0.25 *μ*m x 0.25 *μ*m, whose position may be externally controlled by MEMS. **b,** Simulated (at $\lambda_0 = \lambda_p$ = 10.31 μm) electric field magnitude distribution excited by the QE in the left bubble for displacements: *Δx = 0* (decoupled) and *Δx = 3.5 μm* (coupled). **c,** Decay rate related to coupling $\Gamma_{21}$ (normalized to its free-space counterpart $\Gamma_{21}^{\text{free}}$) and **d,** photonic Lamb shift $\Delta\omega_{21}$ (normalized to its free-space counterpart $\Delta\omega_{21}^{\text{free}}$) as a function of emitter displacement and for different amounts of loss in the ENZ medium ($\varepsilon''$).

# 1 Numerical simulations

The commercially available full-wave electromagnetic simulator software COMSOL Multiphysics®, version 5.0 [S1], was used to compute the dyadic Green's functions and the field distributions displayed in Figs. 1, 2 and 3 of the main text. Specifically, we carried out analysis in the frequency domain solver, where the quantum emitter (QE) was modeled as a point dipole source with dipole moment $\mathbf{p}$ [S2, S3]. The Green's functions and field distributions provided by the numerical solver were employed to compute the following related quantities: effective electrostatic dipole moment, excitation rate, quantum efficiency, decay rate associated to coupling and photonic Lamb shift. First, the effective dipole moment $p_{\text{eff}}$ depicted in Fig. 1c was numerically computed by evaluating the field at the position $(0,0,1.05\,r_0)$, where $r_0$ is the radius of the vacuum bubble containing the QE, and normalizing it respect to that of an electrostatic dipole with the same dipole moment. The excitation rate $\Gamma_{\text{exc}}$ was computed following [S4, S5], i.e., $\Gamma_{\text{exc}} = \Gamma_{\text{rad}}\,D_{\text{rad}}$, where $\Gamma_{\text{rad}}$ is the radiative decay rate and $D_{\text{rad}}$ is the radiation directivity, which were found via the surface integration of the numerically computed fields:

$$\Gamma_{\text{rad}} = \frac{1}{\hbar\omega} \oint_S \mathbf{S}\left(\widehat{\mathbf{r}}\right) \cdot \widehat{\mathbf{n}}\, dS \tag{1}$$

$$D_{\text{rad}}\left(\widehat{\mathbf{r}}\right) = 4\pi r^2 \frac{\widehat{\mathbf{r}} \cdot \mathbf{S}\left(\widehat{\mathbf{r}}\right)}{\hbar\omega \Gamma_{\text{rad}}} \tag{2}$$

with $\mathbf{S}\left(\widehat{\mathbf{r}}\right)$ being the time-average Poynting vector field, $\mathbf{S}\left(\widehat{\mathbf{r}}\right) = 1/2\,\text{Re}\left\{\mathbf{E} \times \mathbf{H}^*\right\}$. The excitation rate was normalized with respect to its free-space counterpart, $\Gamma_{\text{exc}}^{\text{free}} = \Gamma_{\text{rad}}^{\text{free}} D_{\text{rad}}^{\text{free}}$, with $\Gamma_{\text{rad}}^{\text{free}} = \omega^3 |\mathbf{p}|^2 / (12\pi\varepsilon_0 \hbar c^3)$ and $D_{\text{rad}0}^{\text{free}} = 1.5$. The quantum efficiency $\eta_{\text{rad}}$ was computed as the ratio between the radiative and total decay rates [S4, S5]

$$\eta_{\text{rad}} = \frac{\Gamma_{\text{rad}}}{\Gamma_{\text{rad}} + \Gamma_{\text{loss}} + \Gamma_{\text{NR}}} \tag{3}$$

where $\Gamma_{\text{NR}}$ is the intrinsic nonradiative decay rate of the QE and we assume $\Gamma_{\text{NR}} = \Gamma_{\text{rad}}^{\text{free}}$. The nonradiative decay rate in the cavity $\Gamma_{\text{loss}}$ was computed via volume integration of the electric field intensity [S6]

$$\Gamma_{\text{loss}} = \frac{1}{\hbar\omega} \int_V \frac{\omega\varepsilon_0\varepsilon''}{2} |\mathbf{E}|^2\, dV \tag{4}$$

The decay rate associated with coupling and photonic Lamb shift were directly computed from the Green's functions [S7]

$$\Gamma_{21} = \frac{2k_0^2}{\hbar\varepsilon_0} \mathbf{p}_2 \cdot \text{Im}\left\{\overline{\overline{\mathbf{G}}}\left(\mathbf{r}_2, \mathbf{r}_1\right)\right\} \cdot \mathbf{p}_1 \tag{5}$$

$$\triangle\omega_{21} = -\frac{k_0^2}{\hbar\varepsilon_0} \mathbf{p}_2 \cdot \text{Re}\left\{\overline{\overline{\mathbf{G}}}\left(\mathbf{r}_2, \mathbf{r}_1\right)\right\} \cdot \mathbf{p}_1 \tag{6}$$

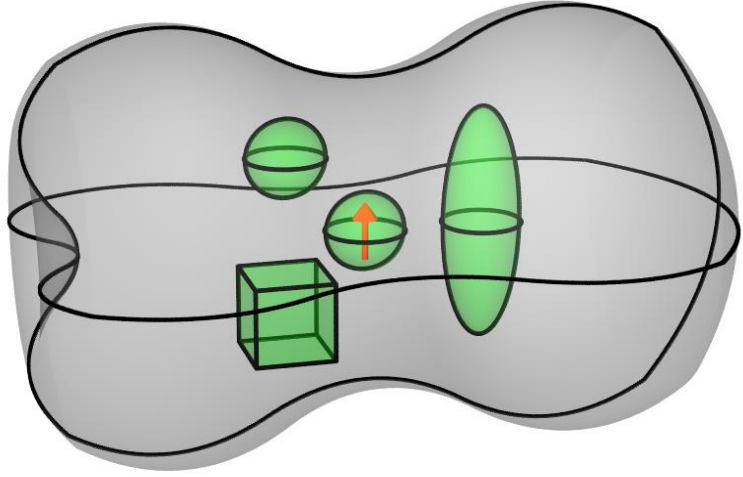

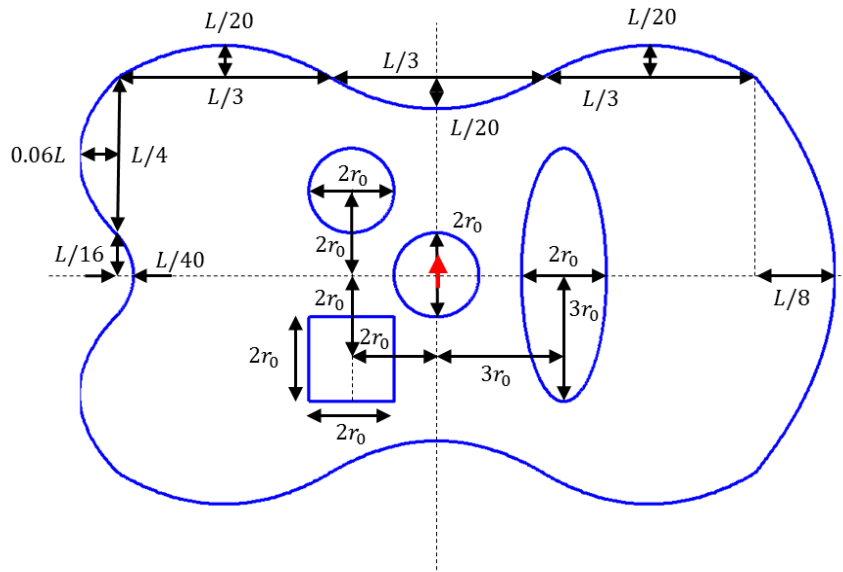

**Fig. S1.** Sketch (top) and dimensions (bottom) of the system studied in Fig. 2 of the main text. The cavity consists of an open epsilon-near-zero volume (shown as grey background) containing a few vacuum bubbles (shown in green) and a quantum emitter (shown as red arrow). The blue curves correspond to the second order polynomials $f(u) = c_0 + c_1 u + c_2 u^2$ that fit to the specified dimensions.

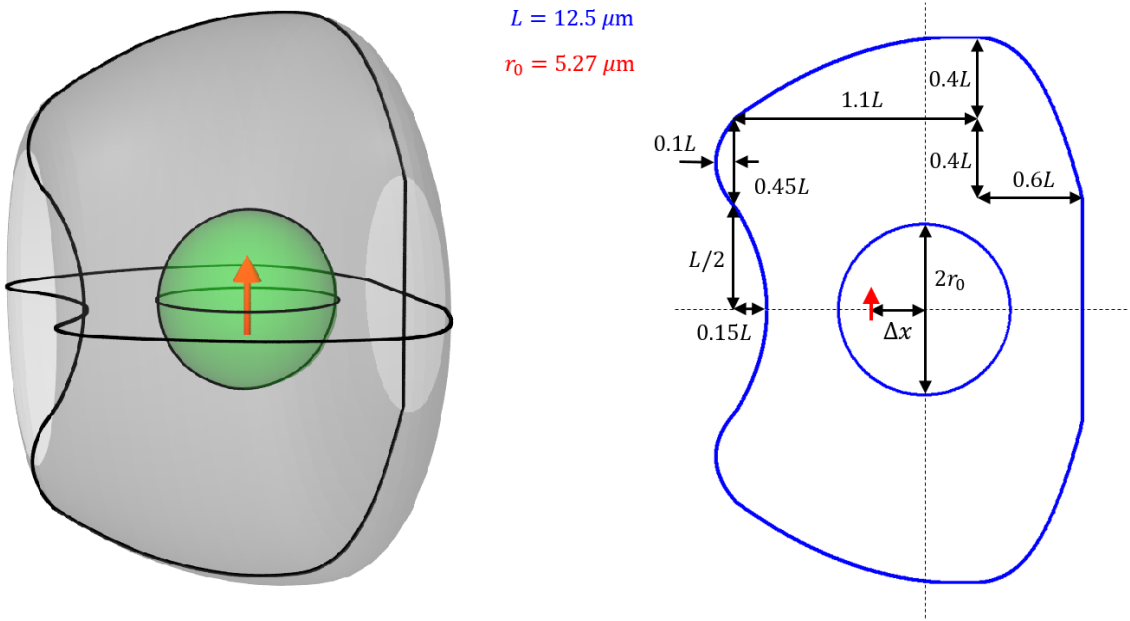

**Fig. S2.** Sketch (left) and dimensions (right) of the system studied in Fig. 3 of the main text. The cavity consists of an open epsilon-near-zero volume (shown as grey background) containing a spherical vacuum bubble (shown in green) and a quantum emitter (shown as red arrow). The blue curves correspond to the second order polynomials $f(u) = c_0 + c_1 u + c_2 u^2$ that fit to the specified dimensions.

$L = 15\ \mu m$

$r_0 = 1.505\ \mu m$

$t = 0.25\ \mu m$

**Fig. S3.** Sketch (top) and dimensions (bottom) of the system studied in Fig. 4 of the main text. The cavity consists of an open epsilon-near-zero volume (shown as grey background) containing two silicon (Si) spherical bubbles (shown in green) and a quantum emitter (shown as red arrow). The cavity is also pierced by a Si rod (shown in green). The blue curves correspond to the second order polynomials $f(u) = c_0 + c_1 u + c_2 u^2$ that fit to the specified dimensions.

# 3 Quantum emitter contained in a vacuum bubble immersed within epsilon-near-zero media

In this section we derive analytical expressions for the fields excited by a quantum emitter (QE) contained in a vacuum bubble immersed within an epsilon-near-zero (ENZ) background medium. To this end, we first introduce the general solution to Maxwell equations for the electromagnetic field excited by an arbitrary distribution of sources contained in a vacuum bubble immersed within an unbounded medium, and then we observe the solution of a QE in ENZ media as a limiting case. Time-harmonic field expressions $\exp(-i\omega t)$ are assumed and omitted hereafter.

## 3.1 General solution

Let us then consider a distribution of sources $\mathbf{J}(\mathbf{r})$ immersed within a background medium characterized by relative permittivity $\varepsilon$, propagation constant $k$ and intrinsic medium impedance $\eta$. In order to insulate the sources from the background medium, we assume that they are contained within a vacuum spherical bubble of radius $r_0$ (see Fig. S4).

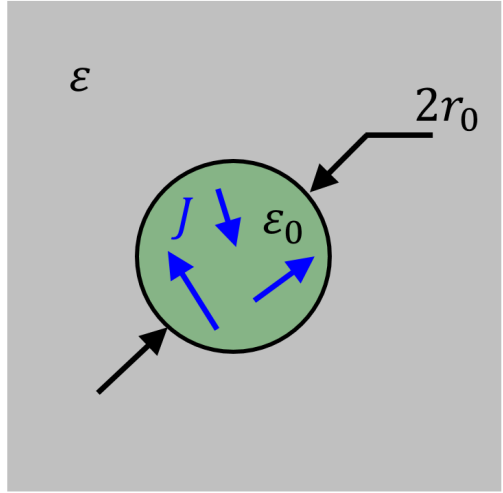

**Fig. S4.** Sketch of a distribution of currents $\mathbf{J}(\mathbf{r})$ immersed in a background medium of relative permittivity $\varepsilon$ but insulated from it by a vacuum sphere of radius $r_0$.

Without loss of generality, the internal and external fields to the bubble can be written as a multipolar decomposition of Tesseral harmonics [S6]

$$\mathbf{E}^{int} = \sum_{\{q\}} \left[ i\, a_{nm}^{lTM} \mathbf{N}_{nm}^l - a_{nm}^{lTE} \mathbf{M}_{nm}^l \right] \tag{7}$$

$$\mathbf{H}^{int} = \frac{1}{\eta_0} \sum_{\{q\}} \left[ a_{nm}^{lTM} \mathbf{M}_{nm}^l + i\, a_{nm}^{lTE} \mathbf{N}_{nm}^l \right] \tag{8}$$

$$\mathbf{E}^{ext} = \sum_{\{q\}} \left[ i\, a_{nm}^{lTM} b_{nm}^{lTM} \mathbf{N}_{nm}^l - a_{nm}^{lTE} b_{nm}^{lTE} \mathbf{M}_{nm}^l \right] \tag{9}$$

$$\mathbf{H}^{ext} = \frac{1}{\eta} \sum_{\{q\}} [a_{nm}^{lTM} b_{nm}^{lTM} \mathbf{M}_{nm}^l + i\, a_{nm}^{lTE} b_{nm}^{lTE} \mathbf{N}_{nm}^l] \qquad (10)$$

where $\{q\} = \{n, m, l\}$ is a multi-index defined so that the sum runs over all spherical multipoles:

$$\sum_{\{q\}} = \sum_{n=1}^{\infty} \sum_{m=0}^{n} \sum_{l=e,o} \qquad (11)$$

$\mathbf{N}_{nm}^l$ and $\mathbf{M}_{nm}^l$ are the Stratton vector fields, which are defined as follows

$$\mathbf{M}_{nm}^l(\mathbf{r}) = \frac{1}{k} \nabla \times \left\{ \widehat{B}_n(kr) T_{nm}^l(\widehat{\mathbf{r}}) \widehat{\mathbf{r}} \right\} \qquad (12)$$

$$\mathbf{N}_{nm}^l(\mathbf{r}) = \frac{1}{k} \nabla \times \mathbf{M}_{nm}^l(\mathbf{r}) \qquad (13)$$

where the angular variation is defined by the Tesseral harmonics

$$T_{nm}^l(\widehat{\mathbf{r}}) = P_n^m(\cos\theta)(\delta_{le}\cos m\phi + \delta_{lo}\sin m\phi) \qquad (14)$$

The functions $\widehat{B}_n(x)$ are linear combinations of Schekunoff form of Bessel spherical functions, i.e., $\widehat{B}_n(x) = \sqrt{\pi x/2}\, B_{n+1/2}(x)$ with $B_n(x)$ being any linear combination of the usual cylindrical Bessel functions of order $n$ [S6]. In our case we select

$$\widehat{B}_n(x) = \widehat{H}_n(k_0 r) + c_{nm}^{lTZ} \widehat{J}_n(k_0 r) \;\; r \leq r_0 \qquad (15)$$

$$\widehat{B}_n(x) = \widehat{H}_n(kr) \;\; r > r_0 \qquad (16)$$

in order to describe the fields induced by the sources and cavity modes within the bubble, whereas outgoing waves outside the bubble. On the one hand, the coefficients $\{a_{nm}^{lTM}, a_{nm}^{lTE}\}$ are defined by the source properties. On the other hand, the external $b_{nm}^{lTM}$ and internal $c_{nm}^{lTM}$ field coefficients are found by solving the boundary value problem imposed by the continuity of the tangential fields on the surface of the vacuum bubble. This exercise leads to the following solutions:

$$b_{nm}^{lTM} = i\frac{\eta k}{k_0} \left[ \eta \widehat{J}_n(k_0 r_0) \widehat{H}_n'(kr_0) - \eta_0 \widehat{J}_n'(k_0 r_0) \widehat{H}_n(kr_0) \right]^{-1} \qquad (17)$$

$$c_{nm}^{lTM} = \frac{\eta_0 \widehat{H}_n'(k_0 r_0) \widehat{H}_n(kr_0) - \eta \widehat{H}_n(k_0 r_0) \widehat{H}_n'(kr_0)}{\eta \widehat{J}_n(k_0 r_0) \widehat{H}_n'(kr_0) - \eta_0 \widehat{J}_n'(k_0 r_0) \widehat{H}_n(kr_0)} \qquad (18)$$

$$b_{nm}^{lTE} = -i\,\eta_0 \frac{k}{k_0} \left[ \eta \widehat{J}_n'(k_0 r_0) \widehat{H}_n(kr_0) - \eta_0 \widehat{J}_n(k_0 r_0) \widehat{H}_n'(kr_0) \right]^{-1} \qquad (19)$$

$$c_{nm}^{lTE} = \frac{\eta_0 \widehat{H}_n(k_0 r_0) \widehat{H}_n'(kr_0) - \eta \widehat{H}_n'(k_0 r_0) \widehat{H}_n(kr_0)}{\eta \widehat{J}_n'(k_0 r_0) \widehat{H}_n(kr_0) - \eta_0 \widehat{J}_n(k_0 r_0) \widehat{H}_n'(kr_0)} \qquad (20)$$

## 3.2 External fields excited by an insulated QE immersed in an ENZ background medium

The above set of equations/coefficients represents the general solution to the problem. The fields excited by a quantum emitter immersed in an ENZ background medium can be found as a limiting case of such a solution. First, due to the symmetry of the problem the fields excited by a QE characterized by dipole moment $\mathbf{p} = \widehat{\mathbf{z}} p$ correpond to those of the $n=1$, $m=0$, $l=e$ multipole, with source coefficient $a_{10}^{eTM} = \omega \eta_0 k_0^2 / (4\pi)\, p$. Second, the impact of the ENZ background medium can be evaluated by taking the limits $\eta \to \infty$, $k \to 0$. In this manner, the external fields can be asymptotically written as follows:

$$\mathbf{E}^{ext} = \frac{\widehat{\mathbf{r}}\, 2\cos\theta + \widehat{\boldsymbol{\theta}}\sin\theta}{4\pi\varepsilon_0 r^3} \left[ \frac{(k_0 r_0)^2}{\widehat{J}_1(k_0 r_0)} p \right] \tag{21}$$

$$\mathbf{H}^{ext} = 0 \tag{22}$$

By comparing these fields with those excited by an electrostatic dipole it is clear that the spatial distribution of the fields excited by an insulated QE immersed in a ENZ medium corresponds to those of an electrostatic dipole (even though the dipole is dynamically oscillating with time with radian frequency $\omega$), with effective dipole moment

$$p_{\text{eff}} = \frac{(k_0 r_0)^2}{\widehat{J}_1(k_0 r_0)} p \tag{23}$$

Note that according to (23) the effective electrostatic dipole moment is resonant at $\widehat{J}_1(k_0 r_0) = 0$, and it asymptotically converges to $p_{\text{eff}} \simeq 3p$ for $k_0 r_0 \ll 1$.

## 3.3 Solution to the equivalent quasistatic problem

Here we demonstrate that the $k_0 r_0 \ll 1$ limit of the general solution is consistent with the quasistatic solution to the problem. To this end, note that the quasi-static fields internal and external to the sphere containing the dipole can be written as

$$\mathbf{E}^{ext} = C \frac{\widehat{\mathbf{r}}\, 2\cos\theta + \widehat{\boldsymbol{\theta}}\sin\theta}{r^3} \tag{24}$$

$$\mathbf{E}^{int} = A \frac{\widehat{\mathbf{r}}\, 2\cos\theta + \widehat{\boldsymbol{\theta}}\sin\theta}{r^3} + B \left( \widehat{\mathbf{r}}\cos\theta - \widehat{\boldsymbol{\theta}}\sin\theta \right) \tag{25}$$

Solving the boundary value problem at $r = r_0$ we determine the value of the $B$ and $C$ coefficients as a function of the source coefficient $A$. These are given by

$$C = A \frac{3\varepsilon_i}{\varepsilon_i + 2\varepsilon_h} \tag{26}$$

$$B = 2 \frac{A}{r_0^3} \frac{\varepsilon_i + \varepsilon_h}{\varepsilon_i + 2\varepsilon_h} \tag{27}$$

Therefore, it is clear than in the ENZ limit ($\varepsilon_h \to 0$) we find $C \to 3A$. Consequently, the external dipole is effectively three times larger than the internal dipole, in agreement with the full time-harmonic analysis derived in the previous section.

## 3.4 Effective electrostatic dipole at resonance

The expression for the effective electrostatic dipole moment (23) exhibits a resonance at $\widehat{J}_1(k_0 r_0) = 0$, where, in the absence of dissipation losses, $p_{\text{eff}}$ becomes arbitrarily large. Here we introduce a correction expression for a finite amount of loss, characterized by the imaginary part of the relative permittivity $\varepsilon''$ of the background medium. To this end, we explicitely evaluate $\widehat{J}_1(k_0 r_0) = 0$ in the external field coefficient (17) and take the $\varepsilon \to 0$ limit. In doing so, the external field coefficient can be written as follows

$$b_{10}^{eTM} \simeq \frac{\eta k^2}{k_0} \frac{r_0}{\eta_0 \widehat{J}'_n(k_0 r_0)} \qquad (28)$$

Consequently, the external field at resonance is given by

$$\mathbf{E}_{TM}^{ext} \simeq \frac{2\cos\theta \widehat{\mathbf{r}} + \widehat{\boldsymbol{\theta}}\sin\theta}{4\pi\varepsilon_0 r^3} \left[ \frac{-i}{\varepsilon''} \frac{k_0 r_0}{\widehat{J}'_1(k_0 r_0)} p \right] \qquad (29)$$

Thus, the effective electrostatic dipole at resonance can be approximated as follows:

$$p_{\text{eff}} \simeq -\frac{i}{\varepsilon''} \frac{k_0 r_0}{\widehat{J}'_1(k_0 r_0)} p \simeq -i \frac{4.6}{\varepsilon''_r} p \qquad (30)$$

## 3.5 Magnetic dipole resonance

Here we identify the radii of the vaccum spherical bubbles for which the magnetic dipole mode is at resonance. Inspecting the coefficients (19)-(20) it is clear that the magnetic dipole resonance of the general bubble-unbounded media system appears at the solutions of the following characteristic equation:

$$\eta \widehat{J}'_1(k_0 r_0) \widehat{H}_1(k r_0) - \eta_0 \widehat{J}_1(k_0 r_0) \widehat{H}'_1(k r_0) = 0 \qquad (31)$$

Moreover, in the ENZ ($\varepsilon \to 0$) limit such a characteristic equation reduces to

$$\widehat{J}'_1(k_0 r_0) + \frac{\widehat{J}_1(k_0 r_0)}{k_0 r_0} = 0 \qquad (32)$$

The figure below depicts the l.h.s. of the characteristic equation as a function of $k_0 r_0$. It is apparent from the figure that the magnetic dipole resonance takes place approximately at $k_0 r_0 \simeq 3.14$. For a vaccum bubble operating at $\lambda = 10.32\,\mu$m this value corresponds to a radius of $r_0 = 5.17\,\mu$m, whereas for a silicon ($\varepsilon_r = 11.7$) bubble it corresponds to a radius of $r_0 = 1.507\,\mu$m.

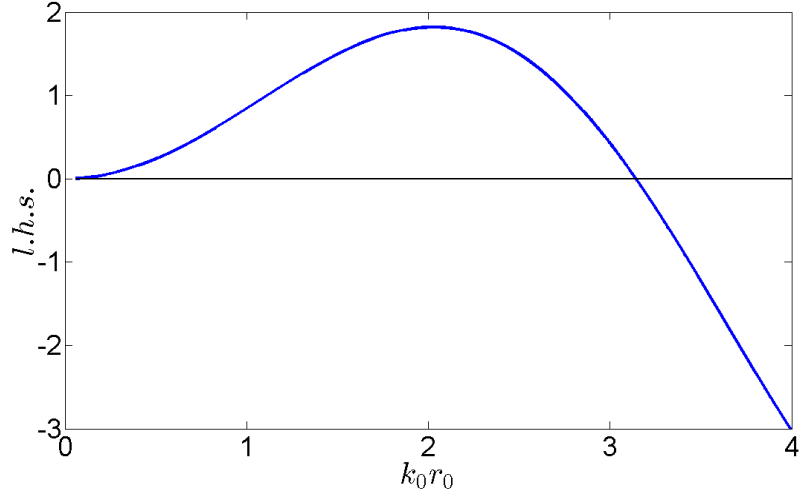

**Fig. S5.** Left hand side (l.h.s.) of the characteristic equation (32) of the magnetic dipole resonance as a function the spherical bubble electrical size $k_0 r_0$. The resonance is excited when the l.h.s. equals zero.

## 3.6 QEs shifted from the origin of the coordinates

As the position of the QE is shifted from the origin of the coordinates, the fields excited by it can be written as a series of multipole sources centered at the origin of the coordinates. Specifically, the source coefficients for a QE with dipole moment $\mathbf{p}$ placed at the position $\mathbf{r}'$ are given by [S8]

$$a_{nm}^{lTM} = -\omega \frac{\eta_0 k_0^2}{f_{nm}} \mathbf{p} \cdot \mathbf{N}_{nm}^l(\mathbf{r}') \tag{33}$$

$$a_{nm}^{lTE} = i\omega \frac{\eta_0 k_0^2}{f_{nm}} \mathbf{p} \cdot \mathbf{M}_{nm}^l(\mathbf{r}') \tag{34}$$

with

$$f_{nm} = (1 + \delta_{m0}) \frac{2\pi n(n+1)}{2n+1} \frac{(n+m)!}{(n-m)!} \tag{35}$$

Next, for a dipole positioned on the x-axis an oriented along $\hat{\mathbf{z}}$ the coefficients reduce to

$$a_{nm}^{lTM} = \frac{\widehat{J}'_n(k_0 \triangle x)}{k_0 \triangle x} p \, \psi_{nml}^{TM} \tag{36}$$

$$a_{nm}^{lTE} = \frac{\widehat{J}_n(k_0 \triangle x)}{k_0 \triangle x} p \, \psi_{nml}^{TE} \tag{37}$$

with

$$\psi_{nml}^{TM} = -\omega \frac{\eta_0 k_0^2}{f_{nm}} \delta_{le} P_n^{m\,\prime}(0) \tag{38}$$

$$\psi_{nml}^{TE} = i\omega \frac{\eta_0 k_0^2}{f_{nm}} \delta_{lo} m \, P_n^m(0) \tag{39}$$

It is thus clear that the electric and magnetic dipole excitations oscillate following $\widehat{J}_1'(k_0 \triangle x)/(k_0 \triangle x)$ and $\widehat{J}_1(k_0 \triangle x)/(k_0 \triangle x)$, respectively, as the QE is shifted along the x-axis. Both functions are depicted in Fig. S6. It is apparent from the figure that the optimal position for the excitation of the magnetic dipole response corresponds to the displacement $k_0 \triangle x \simeq 2.08$, or, equivalently, $\triangle x \simeq 2.08\lambda/(2\pi\sqrt{\varepsilon_i})$. Subsequently, at $\lambda = 10.32\,\mu$m the optimal displacements equal $\triangle x \simeq 3.42\,\mu$m in vacuum (i.e., $\varepsilon_i = 1$) and $\triangle x \simeq 1.00\,\mu$m in silicon (i.e., $\varepsilon_i = 11.7$).

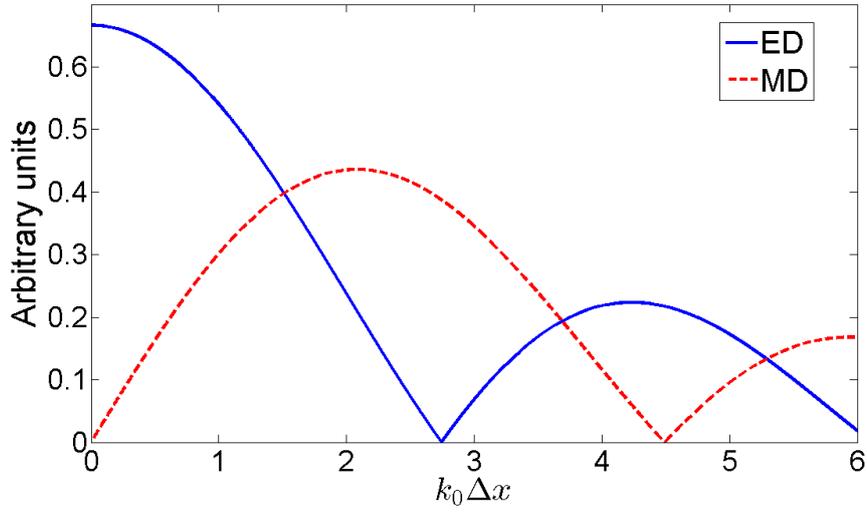

**Fig. S6.** Magnitude of the electric dipole (ED) and magnetic dipole (MD) coefficients (in arbitrary units) as a function of the quantum emitter position displacement $k_0 \triangle x$.

# References

[S1] COMSOL Multiphysics® (www.comsol.com).

[S2] L. Novotny and B. Hecht. Principles of Nano-Optics. Cambridge University Press, 2006.

[S3] W. Demtroder. Atoms, Molecules and Photons: An Introduction to Atomic-, Molecular- and Quantum Physics. Springer, 2006.

[S4] L. Novotny and N. van Hulst, "Antennas for light," Nat. Photon., vol. 5, no. 2, pp. 83–90, Feb. 2011.

[S5] P. Bharadwaj, B. Deutsch, and L. Novotny, "Optical Antennas," Adv. Opt. Photonics, vol. 1, no. 3, p. 438, Aug. 2009.

[S6] R. F. Harrington, Time-Harmonic Electromagnetic Fields (McGraw-Hill, New York, 1961).

[S7] P. A. Huidobro, A. Y. Nikitin, C. González-Ballestero, L. Martín-Moreno, and F. J. García-Vidal, "Superradiance mediated by graphene surface plasmons," Phys. Rev. B, vol. 85, no. 15, p. 155438, 2012.

[S8] I. Liberal, I. Ederra, R. Gonzalo, and R. W. Ziolkowski, "A multipolar analysis of near-field absorption and scattering processes," IEEE Trans. Antennas Propag. , vol. 61, no. 10, pp. 5184–5199, Oct. 2013.